# Phase transition in ultrathin magnetic films with long-range interactions: Monte Carlo simulation of the anisotropic Heisenberg model

M. Rapini,* R. A. Dias,† and B. V. Costa‡

*Departamento de Física, Laboratório de Simulação, ICEX, UFMG 30123-970 Belo Horizonte, MG, Brazil*

Ultrathin magnetic films can be modeled as an anisotropic Heisenberg model with long-range dipolar interactions. It is believed that the phase diagram presents three phases: An ordered ferromagnetic phase (I), a phase characterized by a change from out-of-plane to in-plane in the magnetization (II), and a high-temperature paramagnetic phase (III). It is claimed that the border lines from phase I to III and II to III are of second order and from I to II is first order. In the present work we have performed a very careful Monte Carlo simulation of the model. Our results strongly support that the line separating phases II and III is of the BKT type.

## I. INTRODUCTION

Since the late 1980s there has being an increasing interest in ultrathin magnetic films.[1–6] This interest is mainly associated with the development of magnetic-nonmagnetic multilayers for the purpose of giant magnetoresistence applications.[7] In addition, experiments on epitaxial magnetic layers have shown that a huge variety of complex structures can develop in the system.[8–11] Rich magnetic domain structures like stripes, chevrons, labyrinths, and bubbles associated with the competition between dipolar long-range interactions and strong anisotropies perpendicular to the plane of the film were observed experimentally. A lot of theoretical work has been done on the morphology and stability of these magnetic structures.[12–14] Beside that, it has been observed the existence of a switching transition from perpendicular to in-plane ordering at low but finite temperature:[15–18] at low temperature the film magnetization is perpendicular to the film surface; as temperature rises the magnetization flips to an in-plane configuration. Eventually the out-of-plane and the in-plane magnetization become zero.[19]

The general Hamiltonian describing a prototype for an ultrathin magnetic film assumed to lay in the $xy$ plane is[17]

$$H = -J\sum_{\langle ij\rangle} \vec{S}_i \cdot \vec{S}_j - A\sum_i S_i^{z2}$$
$$+ D\sum_{\langle ij\rangle} \left[ \frac{\vec{S}_i \cdot \vec{S}_j}{r_{ij}^3} - 3\frac{(\vec{S}_i \cdot \vec{r}_{ij})(\vec{S}_j \cdot \vec{r}_{ij})}{r_{ij}^5} \right]. \quad (1)$$

Here $J$ is an exchange interaction, which is assumed to be nonzero only for nearest-neighbor interaction, $D$ is the dipolar coupling parameter, $A$ is a single-ion anisotropy, and $\vec{r}_{ij} = \vec{r}_j - \vec{r}_i$, where $\vec{r}_i$ stands for lattice vectors. The structures developed in the system depend on the sample geometry and size. Several situations are discussed in Ref. 14 and citations therein.

Although the structures developed in the system are well known, the phase diagram of the model is still being studied. There are several possibilities since we can combine the parameters in many ways. We want to analyze the case $J > 0$ in some interesting situations. A more detailed analysis covering the entire space of parameters is under consideration.

(1) Case $D=0$. For $D=0$ we recover the two-dimensional (2D) anisotropic Heisenberg model. The isotropic case $A=0$ is known to present no transition.[21] For $A>0$ the model is in the Ising universality class[20] undergoing an order-disorder phase transition whose critical temperature is approximately[33]

$$T_2 = \frac{2T_3}{\ln(\pi^2 J/A)}, \quad (2)$$

where $T_3$ is the transition temperature of the three-dimensional Heisenberg model $T_3/J \approx 1.30$.

If $A<0$, the model is in the $xy$ universality class. In this case it is known to have a Berezinskii-Kosterlitz-Thouless (BKT) phase transition.[22–25] This is an unusual magnetic-phase transition characterized by the unbinding of pairs of topological excitations named vortex-antivortex.[26–28] A vortex (antivortex) is a topological excitation in which spins on a closed path around the excitation core precess by $2\pi$ ($-2\pi$). Above $T_{BKT}$ the correlation length behaves as $\xi \approx \exp(bt^{-1/2})$, with $t\equiv(T-T_{BKT})/T_{BKT}$ and $\xi\to\infty$ below $T_{BKT}$.

(2) Case $D\neq 0$. In this case, there is a competition between the dipolar and the anisotropic terms. If $D$ is small

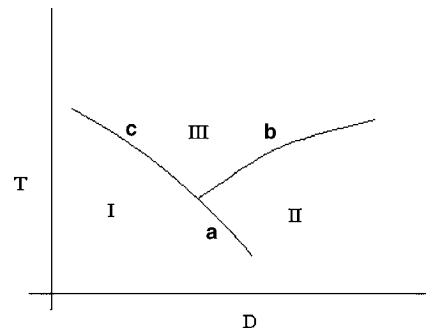

FIG. 1. A sketch of the phase diagram for the model [Eq. (1)]. Phase I corresponds to an out-of-plane magnetization, phase II has in-plane magnetization, and phase III is paramagnetic. The border line between phase I and phase II is believed to be of first order and from regions I and II to III to be both of second order.

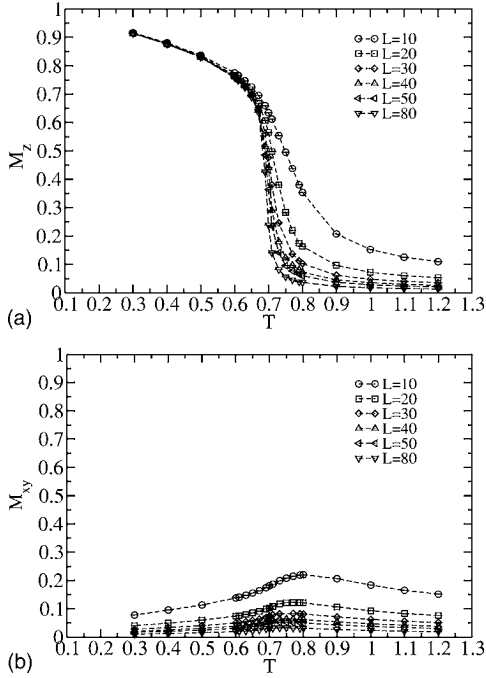

FIG. 2. Out-of-plane (a) and in-plane (b) magnetization for $D=0.1$. The ground state is ferromagnetic. There is no in-plane spontaneous magnetization.

compared to $A$ we can expect the system to have an Ising behavior. If $D$ is not too small we can expect a transition of the spins from out-of-plane to in-plane configuration.[17] For large enough $D$ out-of-plane configurations become unstable such that, the system lowers its energy by turning the spins into an in-plane anti-ferromagnetic arrangement. For the planar $xy$ model with pure dipolar interactions, the system orders at $T_c = 1.39 \pm 0.05$ (Ref. [34]) where temperature is in units of $JS^2/k_B$ and $k_B$ is the Boltzmann constant.

Earlier works on this model, which discuss the phase diagram, were mostly done using renormalization group approach and numerical Monte Carlo simulation.[17,19,26] They agree between themselves in the main features. The phase diagram for fixed $A$ and $J=1$ is schematically shown in Fig. 1 in the space $(D,T)$. From Monte Carlo (MC) results it is found that there are three regions labeled in Fig. 1 as I, II, and III. Phase I corresponds to an out-of-plane magnetization, phase II has in-plane magnetization, and phase III is paramagnetic. The border line between phase I to phase II is believed to be of first order and from regions I and II to III are both second order.

Although the different results agree between themselves about the character of the different regions, much care has to be taken because they were obtained by using a cutoff $r_c$ in the dipolar term. The long-range character of the potential is lost. As a consequence, it will not be surprising if a different phase line emerges coming from region II to region III.

In this work we use MC simulations to investigate the model defined by Eq. (1). We use a cutoff $r_c$ in the dipolar interaction. Our results strongly suggest that the transition between regions II and III is in the BKT universality class, instead of second order, as found in earlier works.

## II. SIMULATION BACKGROUND

Our simulations are done in a square lattice of volume $L \times L$ ($L=10, 20, 30, 40, 50, 80$) with periodic boundary conditions. We use the Monte Carlo method with the Metropolis algorithm.[27,29–31] To treat the dipole term we use a cutoff $r_c = 5a$, where $a$ is the lattice spacing, as suggested in the work of Santamaria and co-workers.[17]

We have performed the simulations for temperatures in the range $0.3 \leq T \leq 1.2$ at intervals of $\Delta T = 0.1$. When necessary this temperature interval is reduced to $\Delta T = 0.01$. For every temperature the first $5 \times 10^5$ MC steps per spin were used to lead the system to equilibrium. The next $10^6$ configurations were used to calculate thermal averages of thermody-namical quantities of interest. We have divided these last $10^6$ configurations in 20 bins from which the error bars are estimated from the standard deviation of the averages over these twenty runs. The single-site anisotropy constant was fixed as $A=1.0$ while the $D$ parameter was set to 0.10, 0.15, and 0.20. In this work the energy is measured in units of $JS^2$ and temperature in units of $JS^2/k_B$, where $k_B$ is the Boltzmann constant.

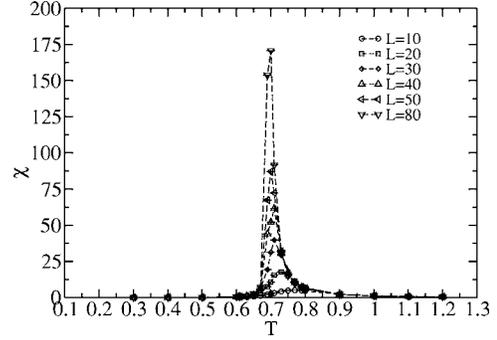

FIG. 4. Out-of-plane susceptibility as a function of temperature for $D=0.1$.

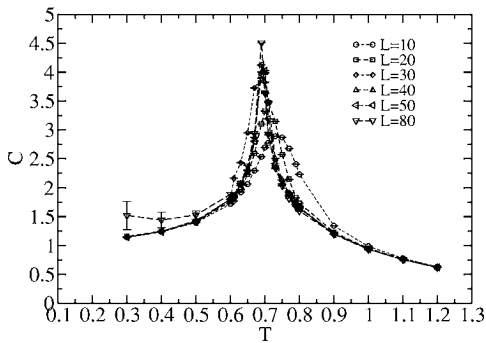

FIG. 3. Specific heat as a function of temperature for $D=0.1$.

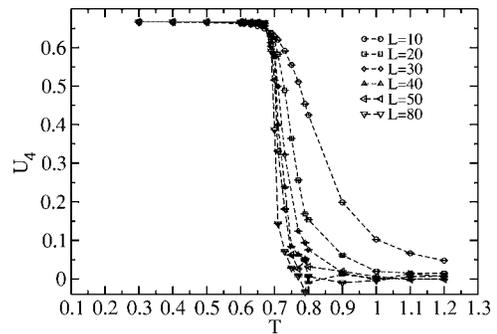

FIG. 5. Binder's cumulant as a function of temperature for $D=0.1$.

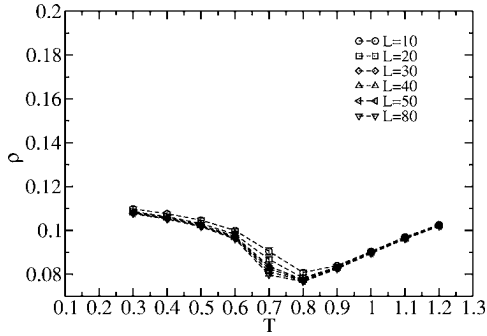

FIG. 6. Vortex density in the $xy$ plane for $D=0.1$.

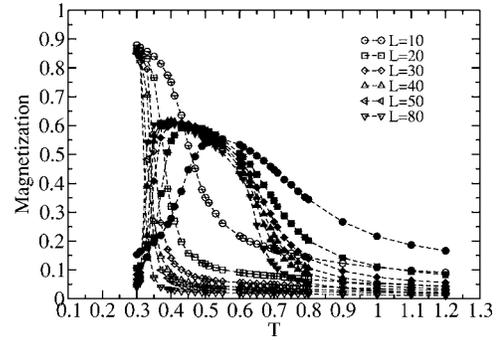

FIG. 7. $M_z$ and $M_{xy}$ (open and full symbols, respectively) for $D=0.15$.

To estimate the transition temperatures we use finite-size-scaling (FSS) analysis to the results of our MC simulations. In the following we summarize the main FSS properties. If the reduced temperature is $t=(T-T_c)/T$, the singular part of the free energy is given by

$$F(L,T) = L^{-(2-\alpha)/\nu}\mathcal{F}(tL^{1/\nu}) \quad (3)$$

for $T$ in the vicinity of the critical temperature and $L$ not too small.

Appropriate differentiation of $F$ yields the various thermodynamic properties. For an order disorder transition exactly at $T_c$ the magnetization $M$, susceptibility $\chi$, and specific heat $C$, behave respectively as[31,32]

$$M \propto L^{-\beta/\nu},$$

$$\chi \propto L^{-\gamma/\nu},$$

$$C \propto L^{-\alpha/\nu}. \quad (4)$$

In addition to these an important quantity is the fourth-order Binder's cumulant

$$U_4 = 1 - \frac{\langle m^4 \rangle}{3\langle m^2 \rangle^2}. \quad (5)$$

where $m$ is the magnetization.

TABLE I. Critical temperature $T_c^L$ of the specific heat $C$, susceptibility $\chi$, and the crosses of the fourth-order Binder's cumulant $U_4$ as a function of the lattice size $L$. Data are for $D=0.10$.

| $L$ | 10 | 20 | 30 | 40 | 50 | 80 |
| --- | --- | --- | --- | --- | --- | --- |
| $C$ | 0.735 | 0.711 | 0.695 | 0.693 | 0.690 | 0.689 |
| $\chi$ | 0.771 | 0.729 | 0.710 | 0.707 | 0.700 | 0.697 |
| $U_4$ | 0.675 | 0.673 | 0.673 | 0.673 | 0.673 | |

For large enough $L$, curves for $U_4(T)$ cross the same point $U^*$ at $T=T_c$. For a BKT transition the quantities defined above behave in a different way. There is no spontaneous magnetization for any finite temperature. The specific heat presents a peak at a temperature that is slightly higher than $T_{BKT}$. Beside that, the peak height does not depend on $L$. Because models presenting a BKT transition have an entire critical region, the curves for $U_4(L)$ just coincide inside that region presenting no crosses at all.

The vortex density is defined as the number of vortices per area. In the simulation, we analyze each plaquette of four sites and if the sum of the difference of the angles between adjacent spins equals $\pm 2\pi$, we have a vortex (antivortex).

Below we present MC results for three typical regions. When not indicated the error bars are smaller than the symbol sizes.

### III. SIMULATION RESULTS

#### A. Case $D=0.1$

For $D=0.1$ we measured the dependence of the out-of-plane magnetization $M_z$ and the in-plane magnetization $M_{xy}$ as a function of temperature for several values of $L$ (see Fig. 2). The figures indicate that in the ground state the system is aligned in the $z$ direction. Approximately at $T \approx 0.70$ the $M_z$ magnetization goes to zero, which gives a rough estimate of the critical temperature. The in-plane magnetization has a small peak close to $T \approx 0.70$. However, the height of the peak diminishes as $L$ grows, in a clear indicative that it is a finite-size artifice.

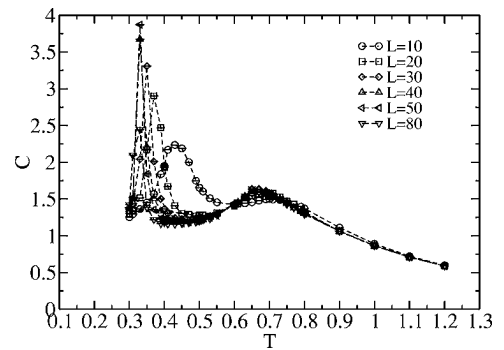

FIG. 8. Specific heat for $D=0.15$.

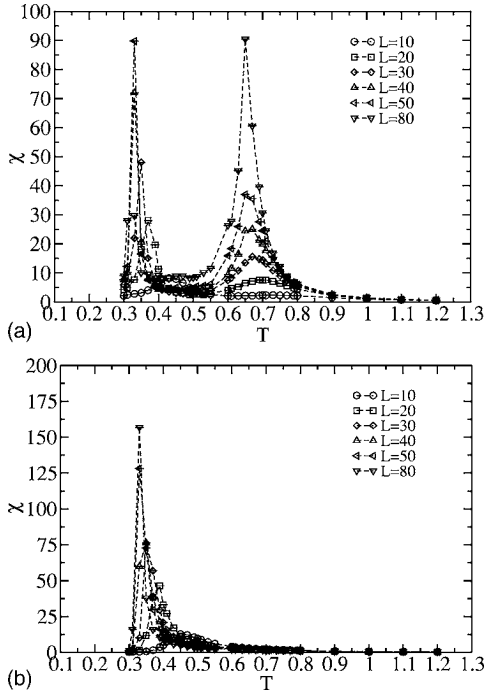

FIG. 9. In-plane (a) and out-of-plane (b) susceptibility for $D=0.15$.

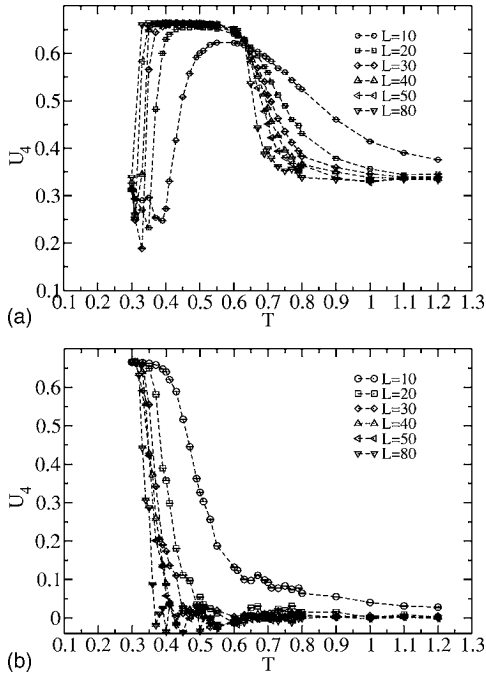

FIG. 10. In-plane (a) and out-of-plane (b) Binder's cumulant as a function of temperature for $D=0.15$. Observe that the in-plane cumulant has a minimum at $T \approx 0.35$ indicating a first-order phase transition. After the minimum the curves do not cross each other having the same behavior (except the spurious case $L=10$). up to $T \approx 0.65$ when they go apart. That is an indication of a BKT phase transition.

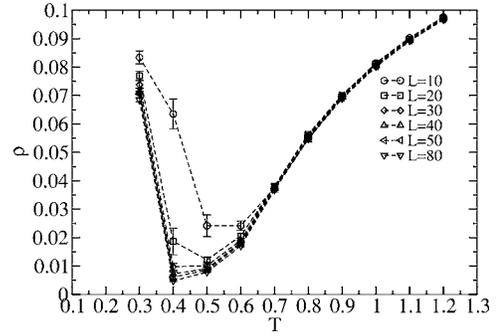

FIG. 11. Vortex density as a function of temperature for $D=0.15$.

The behavior of the specific heat, susceptibility, and Binder's cumulant are shown in Figs. 3, 4, and 5, respectively. The results indicate an order-disorder phase transition in clear agreement with Refs. 5–17 and 19. The vortex density in the $xy$ plane (Fig. 6) has a very shallow minimum near the estimated critical temperature and is almost independent of the lattice size. The growth of the number of vortices when the temperature is decreased is related to the disordering in the plane when the magnetic moments tend to be in the $z$ direction. We have performed a finite-size scaling analysis of the data above by plotting the temperature $T_c^L$ as a function of $1/L$ for the specific heat, the susceptibility, and the crosses of the fourth-order cumulant. The results are shown in Table I. By linear regression we have obtained the critical temperature as $T_c^\infty = 0.682(2)$. An analysis of the maxima of the specific heat $C_{max}$ (see Fig. 18) as a function of the lattice size shows that it behaves as $C_{max} \propto \ln L$, indicating a second-order phase transition. In the phase diagram we crossed the second-order line labeled $c$.

### B. Case $D=0.15$

In this region of the parameters, it was observed a transition from an out-of-plane ordering at low temperatures to an in-plane configuration as described by the magnetization behavior shown in Fig. 7. We show $M_z$ and $M_{xy}$ in the same figure for comparison. The out-of-plane magnetization goes to zero at $T \approx 0.35$ while an in-plane magnetization sets in. This phenomenon has already been reported experimentally[15,16] and it is due to the competition between the easy axis anisotropy and the dipolar interaction. The specific heat curve presents two peaks (see Fig. 8). The peak at low temperature is pronounced and is centered in the temperature in which occurs the rapid decrease of the out-of-plane magnetization $T_1 \approx 0.35$. The second peak appears at $T_2 \approx 0.65$ and seems to be independent of the lattice size.

TABLE II. Critical temperature $T_c^L$ as a function of the linear size $L$ for the susceptibility $\chi_{xy}$ and $D=0.15$.

| $T_c^L$ | 0.729 | 0.698 | 0.678 | 0.670 | 0.650 | 0.638 |
|---|---|---|---|---|---|---|
| $L$ | 10 | 20 | 30 | 40 | 50 | 80 |

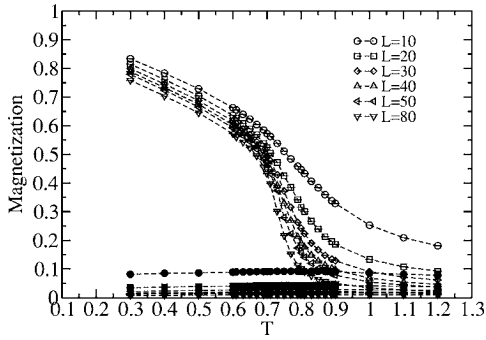

FIG. 12. $M_{xy}$ and $M_z$ (open and full symbols, respectively) for $D=0.2$.

In Fig. 9 we show the in-plane and out-of-plane susceptibilities. The out-of-plane susceptibility presents a single peak close to $T_1 \approx 0.35$. The in-plane susceptibility has a maxima at $T_2 \approx 0.65$ beside the peak at $T_1$, indicating two phase transitions. The Binder's cumulant for the in-plane and out-of-plane magnetization are shown in Figs. 10. Except for the case $L=10$ the curves for different values of the lattice size do not cross each other indicating a BKT transition at $T \approx T_2$. Beside that, the in-plane cumulant has a minimum at $T \approx T_1$, which is a characteristic of a first-order phase transition.[31,32]

The vortex density is shown in Fig. 11. Its behavior is similar to that one shown in Fig. 6. The maxima of the specific heat are shown in Fig. 18 as a function of $L$. It is clear that after a transient behavior it remains constant indicating a BKT transition. A FSS analysis of the susceptibility (see Table II) gives the BKT temperature $T_{BKT}^{\infty}=0.613(5)$. In the phase diagram we crossed the first-order line labeled $a$ ($T_1$) and the line labeled $b$ ($T_2$).

### C. Case $D=0.20$

In Fig. 12 we show the in-plane and out-of-plane magnetization curves for several lattice sizes and $D=0.20$. We observe that as the lattice size $L$ goes from $L=10$ to $L=80$, both magnetizations decrease. It can be inferred that as the system approaches the thermodynamic limit, the net magnetization should be zero. Therefore, the system does not present finite magnetization for any temperature $T \neq 0$. The specific heat (Fig. 13) presents a maximum at $T \approx 0.75$. The curves are for different values of $L$. We observe that the position of the maxima and their heights are not strongly affected by the lattice size, all points falling approximately in the same curve.

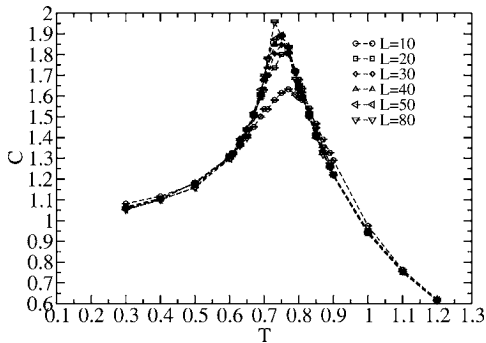

FIG. 13. Specific heat for $D=0.2$. The line is a guide to the eyes.

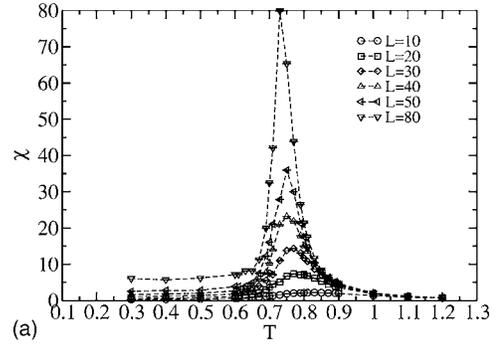

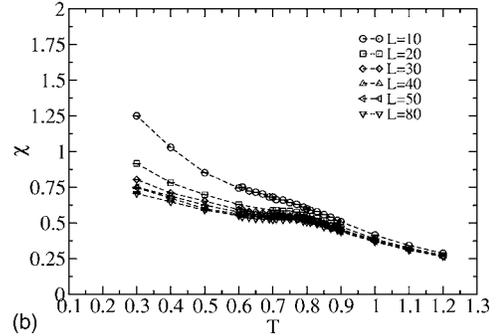

FIG. 14. In-plane (a) and out-of-plane (b) susceptibility for $D=0.2$.

In Fig. 14 we show the in-plane and out-of-plane susceptibilities, respectively. $\chi^{zz}$ does not present any critical behavior. $\chi^{xy}$ presents a peak, which increases with $L$. For the Binder's cumulant (see Fig. 15) there is no unique cross of the curves (except for the $L=10$ curve, which is considered too small to be taken into account). This behavior indicates a BKT transition at $T_{BKT} \approx 0.75$. The vortex density, shown in Fig. 16 is almost independent on the lattice size.

In addition, we did a FSS analysis of the susceptibility (see Table III) and the maxima of the specific heat. The specific heat is shown in Figs. 17 and 18. Its behavior indicates a BKT transition. The analysis of the susceptibility gives $T_{BKT}^{\infty}=0.709(5)$. In the phase diagram we crossed the line labeled $b$. In our results we could not detect any other tran-sition for $D=0.20$, indicating that the line labeled $a$ ends somewhere in between $0.15<D<0.20$ or the crossing at $a$ occurs at a lower temperature ($T<0.30$) outside the range of our simulated data.

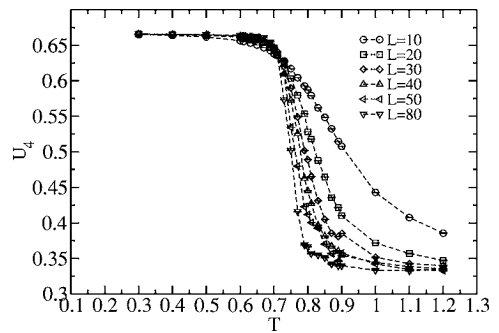

FIG. 15. Fourth-order in-plane cumulant for $D=0.2$.

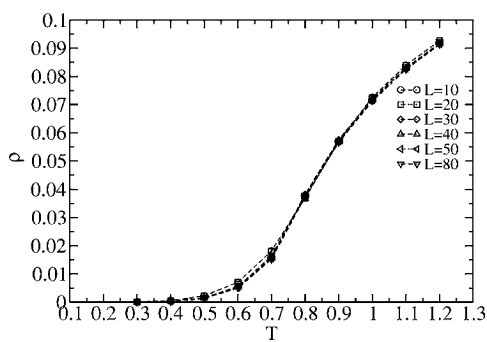

FIG. 16. Vortex density in the $xy$ plane for $D=0.2$.

In a preliminary calculation using a lattice of size $L=40$, we have estimated the value of the multicritical point in the intersection of the $a$, $b$, and $c$ lines around $D=0.14$. Our estimate agrees with the phase diagram obtained by Santamaria and co-workers[17] for $A=2.0$. Their simulations were done on a BCC lattice with (001) surfaces while we used a simple cubic lattice. However, the first layer of these two structures is equivalent.

## IV. CONCLUSIONS

In earlier studies several authors have claimed that the model for ultrathin magnetic films defined by Eq. (1) presents three phases. Referring to Fig. 1 it is believed that the line labeled $a$ is of first order. The lines $b$ and $c$ are of second order. Those results were obtained by introducing a cutoff in the long-range interaction of the Hamiltonian. In the present work we have used a numerical Monte Carlo approach to study the phase diagram of the model for $J=A=1$ and $D=0.10$, 0.15, and 0.20. In order to compare our results to those discussed above we have introduced a cutoff in the long-range dipolar interaction. A finite-size scaling analysis of the magnetization, specific heat, susceptibilities, and Binder's cumulant clearly indicates that the line labeled $a$ is of first order and the line $c$ is of second order in agreement with other results. However, the $b$ line is of BKT type. After analyzing the results obtained, some questions come out:

(1) Is it possible the existence of a limiting range of interaction in the dipolar term beyond which the character of the transition changes from BKT to second order?

(2) How does the line labeled $a$ end in the phase diagram?

(3) What is the character of the intersection point of the three lines in the phase diagram? As the cutoff $r_0$ in the dipolar term is increased, the symmetry of the Hamiltonian is not changed. Therefore, we expect that for larger values of $r_0$, there would be no qualitative changes in our results except when the range of the interaction goes to infinity.

TABLE III. Critical temperature $T_c^L$ as a function of the linear size $L$ for the susceptibility $\chi_{xy}$ and $D=0.20$.

| $T_c^L$ | 0.829 | 0.781 | 0.768 | 0.753 | 0.750 | 0.729 |
|---|---|---|---|---|---|---|
| $L$ | 10 | 20 | 30 | 40 | 50 | 80 |

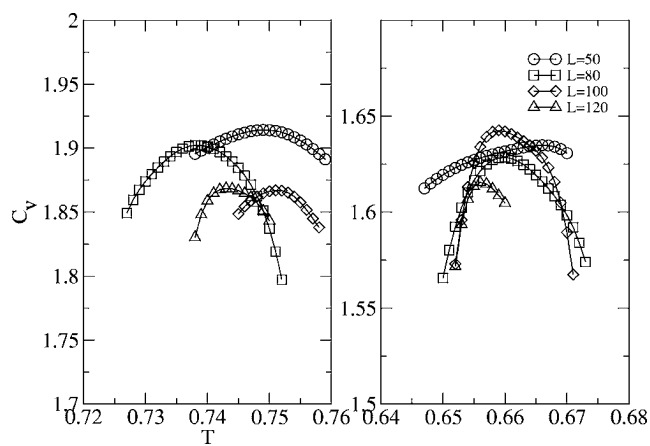

FIG. 17. Specific-heat maxima obtained by using histograms for lattices of sizes $L=50$, 80, 100, and 120 and systems $D=0.20$ (left) and $D=0.15$ (right). Each point is the result of $10^6$ configurations.

However, to respond to question (1) it is necessary for a much more detailed study of the model for several values of the cutoff range $r_c$. In a simulation program we have to be careful in taking larger $r_c$ values since we have to augment the lattice size proportionally to prevent misinterpretations.

In a very preliminary calculation, Rapini *et al.*[18] studied the model with true dipolar long-range interactions by using open boundary conditions and performing the sum without a cutoff. Their results led them to suspect a phase transition of the BKT type involving the unbinding of vortices-antivortices pairs in the model.

In order to estimate the point $(D,T)$ in the phase diagram where the $a$ line ends it would be necessary to study the system for $T<0.3$. Unfortunately, we would not obtain reliable results by using any MC algorithm because for low temperatures the system becomes trapped in a few regions of the phase space for a long time. In the near future, we will address this problem using the histogram technique to respond to questions (2) and (3).

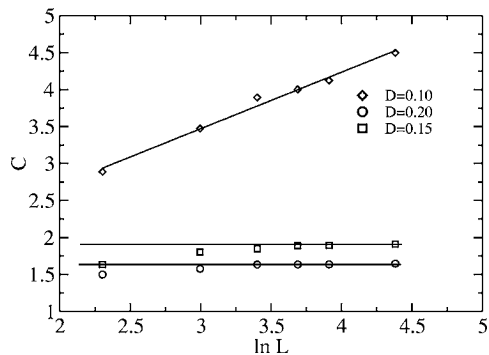

FIG. 18. Maxima of the specific heat as a function of the lattice size. The diamonds are for $D=0.10$, the squares for $D=0.15$, and the circles for $D=0.20$. While in the second-order phase transition the maxima in the specific heat scale as $\ln L$; in the BKT phase transition the finite-size effects are very small.

## ACKNOWLEDGMENTS

Financial support from the Brazilian agencies CNPq, FAPEMIG, and CIAM-0249.0101/03-8 (CNPq) are gratefully acknowledged. Numerical work was done in the LINUX parallel cluster at the Laboratório de Simulação Departamento de Física-UFMG.

---

*Electronic address: mrapini@fisica.ufmg.br
†Electronic address: radias@fisica.ufmg.br
‡Electronic address: bvc@fisica.ufmg.br